\RequirePackage{fix-cm}
\documentclass[smallextended]{svjour3}       
\usepackage[colorlinks,allcolors=blue]{hyperref}
\usepackage{amsbsy,amsmath}
\usepackage{amsfonts}
\usepackage{graphicx}
\usepackage{ulem}
\usepackage{natbib}
\usepackage{breakcites}
\usepackage{lineno}

\title{THE PARAMETRIC OSCILLATOR MODEL FOR THE CASE OF RESONANT ARGUMENT CIRCULATIONS}
\titlerunning{PARAMETRIC MODEL FOR WEAK RESONANT PERTURBATIONS}

\author{Alexey Rosaev \and Eva Pl\'avalov\'a \and Pavel Nesterov}
\authorrunning{A. Rosaev \& E.Plavalova \& P.Nesterov }

\institute{Alexey Rosaev \at Research and Educational Center ``Nonlinear Dynamics'', Yaroslavl State University, Yaroslavl, Russia\\ 
\email{hegem@mail.ru}   
    \and 
Eva Pl\'avalov\'a  \at Slovak Academy of Science, Bratislava, Slovakia\\
\email{plavalova@komplet.sk} 
\and
Pavel Nesterov \at Yaroslavl State University, Yaroslavl, Russia
}

\begin{document}

\date{Received: date / Accepted: date}

\maketitle



\begin{abstract}
The goal of this paper is to obtain an approximate solution of the restricted three-body problem in the case of small perturbations in the vicinity of, but not in exact resonance. In this paper, we study the restricted three-body problem known as planetary type (i.e., when the eccentricity of the test particle is small). 

A method of linearizing the equation of motion close to (but not in) resonance is proposed under the assumption of small perturbations. In other words, we study orbits when the resonant argument circles the resonance. In the practically interesting case of resonant perturbations we can restrict our study to a perturbation with a single frequency with the largest amplitude, and reduce the problem to the Mathieu equation. 

The model qualitatively describes the behavior of the perturbation in the vicinity of the resonance. It can be used to estimate the exact position of the resonance and the boundaries between neighboring resonances.

\end{abstract}

\keywords{
Resonances 
\and Orbital elements 
\and Evolution 
\and Approximated solution 
\and Dynamics }


\section{Introduction}

Goldreich paper \cite{G1} was one of the first studies in which the role of resonances in the solar system was analyzed. Since then, numerous papers have focused on this problem, and the resonance dynamics. The resonance dynamic has been studied intensively. We cite, for example, the books of \cite{MD}, \cite{Mo1}, \cite{FM1} and \cite{Ce2}.

Long-term Solar System evolution is profoundly shaped by orbital resonances, which can promote both dynamical instability and long-term stability. A useful overview with simple models that elucidate our understanding of orbital resonance phenomena is given in paper \cite{Ma}.

In general, a resonance can have its own finite width. This finite width can cause an overlap of nearby resonances, which frequently results in the onset of chaos. As B. V. Chirikov predicted in the last century \cite{Chi}, this resonance overlap plays a significant role in the dynamical behavior of asteroids. Right after Chirikov, \cite{W2} featured resonance overlap in relation to the onset of stochastic behavior of the planar circular restricted three-body problem. After that, the concept of resonance overlap was gradually recognized and accepted by the society of solar system dynamics.  
Resonances are closely related to chaos in dynamical systems. In particular, the problem of so-called stable chaotic motion is discussed in \cite{Ts}, \cite{HM2}.

Many multi-planet systems have been discovered in recent years. A considerable fraction of multi-planet systems discovered by the observational surveys of extrasolar planets reside in mild proximity to first-order mean-motion resonances. This stimulated the study of new aspects of the resonance problem such as planetary migration and resonance capture in the frame of the non-restricted three body problem (\cite{BM1}).

There are many papers studying the resonant motion but most of them deal with the libration solution. In this paper we study the case of resonance argument circulation because even the weak resonant perturbation can have a notable effect on small bodies motion.

There are two fundamental theoretical models used to study resonance. Based on the value of the orbital eccentricity of a particle and also its proximity to the exact resonant orbit in a three-body system, the Pendulum Approximation (\cite{W}, \cite{MD}) or the Second Fundamental Model of resonance  (\cite{HL1}) are commonly used to study the motion of that particle near its resonance state.

Many different methods have been applied to study inner resonances in the asteroid belt. A mapping of the phase space onto itself with the same low-order resonance structure as the 3/1 commensurability in the planar elliptic restricted three body problem is derived in paper \cite{W3}. Based on this technique, Wisdom found that the orbital eccentricity of the small body can suddenly increase after a long period of quiescence. \cite{NFM} applied the frequency map analysis to the following dynamical models of the 2/1 asteroidal mean-motion resonance: the planar restricted three body model and the planar restricted four-body model with Saturn. It allowed them to reproduce the chaotic region, which is formed by the overlap of secondary resonances in low eccentricities. 

The stability of some asteroids, in the framework of the restricted three-body problem, has been proven by developing an isoenergetic KAM theorem. (\cite{Ce3}).

Some important results in the study of resonances were obtained by Gallardo. He suggested a numerical method for calculating the resonance strengths and applied it to the main asteroid belt. The location of the libration centers and their dependence on the orbital elements of the resonant orbit are analyzed in (\cite{Ga1},\cite{Ga2}).

A new study of resonance in the Galilean satellite system proves that it is a three-body resonance, and not a combination of simple resonances of mean motion (\cite{La}).

On the other hand, linear equations with periodic coefficients naturally appear in different physical and mathematical problems so they are very common. These equations describe the behavior of different dynamical systems. This study is devoted to their applications in celestial mechanics. 

Parametric equations are considered in many celestial mechanical works. The linear equation with variable coefficients (Hill equation) plays an important role in the development of the theory of the movement of the Moon. The first approach of the lunar orbit is variational curve, which, in the simplest case, can be represented as a circle. 

The parametric equations are briefly mentioned in the book \cite{Mo1}. We can also note the works of \cite{M1},(\citeyear{M2}). The parametric excited Hamiltonian is written in \cite{HM2}. In the above-mentioned author's papers, the application of the parametric equation to resonance \cite{Sz} is noted.  

In this work, we start from the planar restricted three body problem. The main idea of this research is to use a linear equation and to show that, in some cases, it is possible to obtain very interesting results, that also valid for the complete (nonlinear) problem. 

We show that even a single resonance can excite two perturbations with different frequencies. One of these frequencies always has a non-zero value. This may be important when studying the process of a capture in resonance: the transition between the libration and circulation regimes can occur at a finite non-zero frequency.
We further provide new estimates for the widths of unstable zones associated with a set of well-known mean-motion resonances.

Paper consists of four parts: In the first part, we give the main equation and problem setting. In the second part, the simplest but universal way of  deriving the parametric equation is considered. In the third part, our model is applied to the resonance case. In the fourth part the verification of our model with numerical integration is studied.  


\section{Model description}
 \subsection{Main equations} 
 
The cylindrical coordinate frame has several advantages for many applications of planetary dynamics, due to the consideration of the symmetry problem. The main equations for the restricted 3-body problem in the cylindrical frame are \cite{MD}
 
 \begin{equation}	
  {\frac {{d}^{2}{r}}{{dt}^{2}}}\, - {r}\, \left( \! \,
 {\frac {{ d}}{{ dt}}}\,{ \lambda}\, \! 
  \right) ^{2}=-{\frac {{G}\,{M}}{{r^2}}}+{\frac {{ \partial}{U}}{{ \partial}{r}}}\,
  \label{E1}
  \end{equation}
 
 \begin{equation}
 {\frac {{ d}}{{ dt}}}\,\left( \! \,{r}^{2}\,
  {\frac {{ d}}{{ dt}}}\,{ \lambda}\, \!  \right) ={\frac {{ \partial}{U}}{{ \partial}{ \lambda}
 }}\,
  \label{E2}
 \end{equation}


where $r(t)$ is the distance of a small particle from the central mass $M$, $\lambda $ is the longitude of the small particle.  The perturbation function in the absolute frame consists of a direct and an indirect part:

\begin{equation}
{\mathrm{U}(r, \,\lambda , \,t) = {\frac {{G}\,{m}}{\Delta}}} -{\displaystyle \frac {\mathit{Gm}\,r\,\mathrm{cos}(\delta \,
\lambda )}{R^{2}}} 
 \label{E3}
\end{equation}

where $m$ is the mass of the perturbing body; $r, R$ -- distance from the mass center test particle and perturbing body, respectively; $G$ -- gravity constant,  $\delta\lambda$ -- angle between perturbed and perturbing bodies (differential longitude of perturbed body), and:

 \begin{equation}
 {\Delta}=\sqrt {{R}^{2} + {r}^{2} - 2\,{R}\,{r}\,{\rm cos}(\,{ 
 \delta}\,{ \lambda}\,)}.
  \label{E4}
 \end{equation}

 
  
A mean motion resonance between an asteroid and $j-th$ planet occurs,  when is valid $ k/k_j=n_j/n$, where  $k$ and $k_j$ are positive integers, $n_j$ is the mean motion frequency of $j-th$ planet and $n$ is the mean motion frequency of an asteroid.
  
\begin{figure}
      \centering 
        \includegraphics[width=8.7cm]{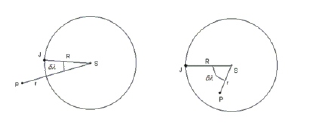}
         \caption{Exterior resonance geometry(left) and interior resonance geometry (right)}
   \label{Pf1}
    \end{figure}

If  $k_j< k$ (the configuration Sun-Planet-Asteroid),  then the resonance is called exterior (outer). {If $k_j> k$ (the configuration Sun-Asteroid-Planet),  then the resonance is called interior (inner) (Fig.  \ref{Pf1}).}
 
\subsection{Parametric equation derivation} 

   In our previous paper (\cite{RP3}), the linear equation with periodic coefficients is derived according to the Laplace method \cite{Su}. However, the classical Laplace calculation method of the perturbation in coordinates aimed to describe motion on the short time interval is mostly along one revolution of the body in its orbit. At each step, the partial solution of non-homogenous equation is necessary to identify, which is not convenient.  As a result,  this method cannot provide a proper picture of evolution over a long time interval.  
For this reason,  we propose an independent method to obtain linear parametric approximation of the restricted three body problem (RTBP).

The principle part of the expansion of the perturbation function may be written using Legendre's polynomials $P_i(\cos\delta\lambda) $.  For the outer source of perturbation,  it is possible to expand the perturbation function in (\ref{E1}) by the power of $\alpha= r/R<1$ \cite{MD}:

\begin{equation}
U_{d} ={\frac {{G}\,{m}}{\Delta}}= \frac {{G}\,{m}}{{R}}\sum\limits_{p=2}^{\infty} (r/{R})^p P_p(\cos(\delta\lambda)).
 \label{E5}
\end{equation}

If the perturbations are assumed to be small, we denote a new variable $x=r-r_0, \  x<<r_0$,  where $r_0$ is either a constant or a known function of time and expand (\ref{E5}) by power x:

\begin{align}
\nonumber
\frac {{ \partial}{U_{d}}}{{ \partial}{r}} = \frac {{G}\,{m}}{{R^2}}\sum\limits_{p=2}^{\infty} p (r_0/{R})^{p-1} P_p(\cos(\delta\lambda))
\\
+\frac {{G}\,{m}}{{R^3}}\sum\limits_{p=2}^{\infty} p (p-1) (r_0/{R})^{p-2}  P_p(\cos(\delta\lambda))x+...
 \label{E6}
\end{align}

Denoting $L=r^2\dot\lambda$ the angular momentum and substituting in equation (\ref{E1}), we obtain: 

\begin{equation}
 {\frac {d ^{2}r}{d t^{2}}}\, - { \frac {L^{2}}{r^{3}}} =
  - {\frac {{GM}}{r^{2 }}}  + {\frac {\partial U}{\partial r}}
   \label{E7}
 \end{equation}
 \begin{equation}
  {\frac {{ d L}}{{ dt}}} ={\frac {{ \partial}{U}}{{ \partial}{ \lambda}
  }}\,
   \label{E8}
  \end{equation}

and expanding:

\begin{equation}
  { \frac {L^{2}}{r^{3}}}- {\frac {{GM}}{r^{2 }}}  ={ \frac {L^{2}}{r_0^{3}}}- {\frac {{GM}}{r_0^{2 }}}-{ \frac {3L^{2}}{r_0^{4}}}x+ {\frac {2{GM}}{r_0^{3 }}}x+...
  \label{E9}
 \end{equation}
 
After that, we can substitute (\ref{E6}) and (\ref{E9}) in (\ref{E1}) and restrict the equation to the first order of $x$,  obtaining the Hill equation for the normal distance from the variation orbit $x$:

\begin{equation}
  \ddot x+\omega^2(t)x=f(t).
   \label{E10}
 \end{equation}
 
 The coefficients in this equation depend on time through longitude:

\begin{align}
\nonumber
  \omega^2 ={ \frac {3L^{2}}{r_0^{4}}}- {\frac {2{GM}}{r_0^{3 }}}-
  \\
  -\frac {{G}\,{m}}{{R^3}}\sum\limits_{p=2}^{\infty} p (p-1) (r_0/{R})^{p-2}  P_p(\cos(\delta\lambda))
   \label{E11}
 \end{align}
 
 \begin{equation}
   f  ={ \frac {L^{2}}{r_0^{3}}}- {\frac {{GM}}{r_0^{2 }}}-\frac {{G}\,{m}}{{R^2}}\sum\limits_{p=2}^{\infty} p (r_0/{R})^{p-1} P_p(\cos(\delta\lambda)).
    \label{E12}
  \end{equation}
 
 If we group the terms with equal $ p\delta\lambda $ in (\ref{E11}),   the main frequency can be performed:
 
  \begin{equation}
    \omega^2  = {\frac {{GM}}{r_0^{3 }}}-\frac {{G}\,{m}}{{R^3}}\sum\limits_{p=2}^{\infty}   b_p(\cos(p\delta\lambda))
     \label{E13}
   \end{equation}
where is possible to calculate $b_p$ numerically.

Now we can take a closer look at the $\delta\lambda$ on time dependence.  In a planar case $\delta\lambda=(n-n_p)t-\varpi$ ,  where $n, n_p $ is the mean motion of the asteroid and the perturbing planet and $\varpi$ is the perihelion longitude of the asteroid. In the first order by eccentricity and zero inclinations (expression (6.90) in \cite{MD}) we have: 

\begin{align}
\nonumber
  \cos(j\delta\lambda)=\cos(j(\lambda-\lambda_p))+
  \\
  \nonumber  
  +je\cos((j-1)\lambda-j\lambda_p+\varpi)+
  \\
  \nonumber  
  +je\cos((j+1)\lambda-j\lambda_p-\varpi)+
  \\
    \nonumber
  +je_p\cos(j\lambda-(j+1)\lambda_p+\varpi_p)-
  \\
  je_p\cos(j\lambda-(j-1)\lambda_p-\varpi_p).
   \label{E14}
 \end{align}
 
 If we are not interested in the short periodic variation of the orbit, we can apply averaging over a fast variable and use it as $f(t)$ in our equations. Therefore, we have: $\delta\lambda=\dot{\varpi}t+\theta $. Here $\theta$ is the value of $\delta\lambda$  at the moment of time $t=0$. The average value of the perturbation function (see expression (6.164) in \cite{MD}) is:
  
 \ \begin{equation}
   f(t)=C_0+ C_1 (e^2+e^2_p) +C_3 e e_p\cos(\dot{\varpi_p}t-\dot{\varpi t})
   \label{E17}
   \end{equation}
 where:
 \begin{equation}
     C_0=1/2b_{1/2}^{(0)}
 \end{equation}
          \begin{align}
          C_1=& \frac{1}{8} \left[ 2\alpha \frac{d}{d \alpha} + \alpha^2 \frac{d^2}{d\alpha^2}  \right] b_{1/2}^{\left( 0 \right) } \left( \alpha \right). 
          \\
          \
         C_3=& \frac{1}{4} \left[ 2 - 2\alpha \frac{d}{d \alpha} + \alpha^2 \frac{d^2}{d\alpha^2}  \right] b_{1/2}^{\left( 1 \right) } \left( \alpha \right). 
          \label{E19}
          \end{align}  
  
    Here $b_{1/2}^{k}$ is the Laplace coefficient, and $\alpha=a/a_p$  is the ratio of the semimajor axis of the asteroid and the planet. 
 
 Note that the expression above has no dependence on the longitude, mean motion and $x$. Therefore,  we can include it in $f(t)$.  After averaging in the absence of resonances,  equation (\ref{E10}), taking into account (\ref{E13}), can be considered as the following:
  
 \begin{equation}
  {\frac {d ^{2}x}{d t^{2}}}\,+ ({ \frac {3L^{2}}{r^{4}}} 
     - {\frac {2{GM}}{r^{3 }}}) x  - \frac {{G}\,{m}}{{R^3}}\sum\limits_{p=2}^{\infty}   b_p(\cos(p\delta\lambda))x+... =f(t).
    \label{E16}
  \end{equation}

   When the perturbations are absent (two body problem),  equation (\ref{E16}) gives the correct solution for a circular orbit.  We can then conclude that $x$ is the radial distance to the circle with a fixed known radius $r_0$.
   
    Additionally, we note that if the orbit of the perturbing planet is circular,  then the variation of the eccentricity has no secular disturbance. Therefore, we can use the eccentricity as well as the semimajor axis to study resonant perturbations.
 
 \subsection{Problem setting and the restrictions of the model}    

Now it is necessary to describe the problem in more detail. First of all, we always consider the planar restricted three body problem, i.e the mass of the small body is equal zero, asteroid, planet (Jupiter, so we assume m/M=0.001) and Sun are orbited in the same fixed plane. 

Usually, the averaged problem is studied (section 3.1). In this case $L=const$  but some resonance terms are necessary to take into account.

To study the parametric instability, we consider non-averaged but linearized problem (i.e. we consider radial perturbation small and take into account only first order of it) in section 3.4. In this case $\partial U/\partial \lambda$  is determined by resonant and short periodic terms.

Here we assume that the Jupiter moves on a circular orbit, however it is not difficult to take into account for planetary eccentricity.

      \begin{figure}
          \centering 
            \includegraphics[width=8.7cm]{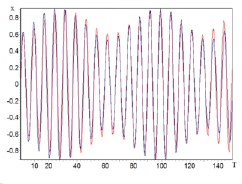}
            \caption{Numerical solution of the Mathieu equation (red) and its approximation (blue)  (in arbitrary units)         }
       \label{Pf3}
        \end{figure}
 
    The presented model is linear, which is its main restriction. The requirement of the linearization is that perturbations are small: $x<<r_0$. In the same time the pendulum model is nonlinear. However, the nonlinearity is important mostly for the libration solution and at the boundary between libration and circulation motion. The formal condition for our linear model is as follows:
\begin{equation}
    \sum\limits_{p=2}^{\infty} p (p-1) (r_0/{R})^{p-2}  P_p(\cos(\delta\lambda))>\frac {{x}}{{2R}}\sum\limits_{p=2}^{\infty} p (p-1)(p-2) (r_0/{R})^{p-3}  P_p(\cos(\delta\lambda)).
\end{equation}
    The other restriction is that we use the Mathieu equation (\ref{E26}) instead of Hill equation (\ref{E10}). Obviously, this is possible for close to resonance orbits. 
      
  \section{Resonant case}
 \subsection{Averaged resonant problem}
 
  To account for the main perturbation in the resonant case we can apply the restrictions on $ \delta\lambda $ :
   \begin{equation}
   \nu t=\delta\lambda=((k+l)n-kn_p)t-l\varpi.
   \label{E25}
   \end{equation}
    Here l, k are integers, and l is the order of resonance.
    Therefore, we can simplify the equation (\ref{E16}) by keeping only the largest terms of perturbation.  As a result,  we have the Mathieu equation instead of the Hill equation: 
   \begin{equation}
     \ddot x+\omega^2_0(1-h_j\cos\nu t)x=f(t)
     \label{E26}
    \end{equation}
    where $h_j$ depends on $b_j$  and $\omega_0=n$ is constant.
    Here index $j$ is the coefficient in expression for cosine in (\ref{E14}). Substituting the resonant term (\ref{E14}) in (\ref{E16}) for the 3:2 resonance we have (j=3):
\begin{equation}
    h_3\cos \nu t=3e b^3_{3/2}\frac {m r_{0}^3}{M R^3}\cos[-2\lambda+3\lambda_p-\varpi].
     \label{E22}
\end{equation}
For the 2:1 resonance we have (j=2):
\begin{equation}
    h_2\cos \nu t=2e b^2_{3/2}\frac {m r_{0}^3}{M R^3}\cos[-\lambda+2\lambda_p-\varpi].
     \label{E27}
\end{equation}
 
    Here $b_{3/2}^{3}$ and $b_{3/2}^{2}$ are the Laplace coefficients. 

\subsection{Properties of Mathieu equation solution}

       
 The asymptotic solution of the Mathieu equation by \cite{BM} contains two frequencies. The solution of equation (\ref{E26}) far from parametric instability is:
     
       \begin{equation}
       x=(c_4\sin(\Lambda t)+c_5\cos(\Lambda t))\cos(\nu/2 t+\theta)
       \label{E31}
       \end{equation}
                 
       \begin{equation}
         \Lambda^2=h_j^2\omega_0^4/(4\nu)-(\omega_0-\nu/2)^2
        \label{E32}
         \end{equation}
    
    where $c_4$, $c_5$, and $\theta$ are constants, which are determined by the initial conditions. 

However, the calculation of these constants is non trivial. For this reason, to study the dependence of the frequency of perturbations, we found a series of numerical solutions of the Mathieu equation with different distances to resonance and different masses of perturbing planet. We approximate the solution (\ref{E26}) in the form:

    \begin{equation}
       x=c_4\cos(\Lambda t+\nu/2t+\theta_1)+c_5\cos(\Lambda t-\nu/2t+\theta_2).
       \label{E31b}
       \end{equation}
For example, for the equation:

\begin{equation}
       \ddot x +0.70 x(1+0.10\cos(1.780128 t))=0.
 \end{equation}
We have an approximate solution (Fig.\ref{Pf3}): 
\begin{equation}
x=0.7\cos(0.839 t+4.8)-0.15\cos(0.935 t+5).
 \end{equation}
 
 A similar behaviour is noted in the evolution of orbital elements of some real small bodies near the resonance (\cite{Ce1}, \cite{BF}). Note that the presence of $f(t)$ on the right side of (\ref{E26}) gives an additional term to the solution with the frequency of the force of perturbation.
 
    \begin{figure}
       \centering 
         \includegraphics[width=8.7cm]{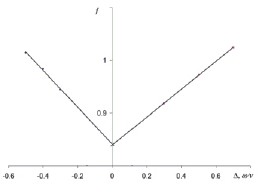}
         \caption {The behavior of the frequency of the approximated solution of the Mathieu equation 
         in dependence of distance to resonance. Here the values of frequency $\nu$ (see equation(\ref{E31})) are on the vertical axis and the distance to resonance (in frequency) is on the horizontal axis.}
    \label{Pf4}
     \end{figure}
   
   We notice that the values of both frequencies decrease as we approach resonance. The highest frequency has $\nu=\omega_0$ as the limit (minimum value), the lowest frequency ($\Lambda$) has a zero minimum value (Fig.\ref{Pf4}). A similar quasi-linear dependence of one of the fundamental frequencies on the distance to the 2:1 resonance is noted in the paper by \cite{La}.
   
\begin{figure}
       \centering 
         \includegraphics[width=8.7cm]{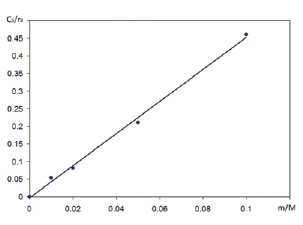}
        \caption {The dependence of the amplitude of the numerical solution of the Mathieu equation on the mass of the perturbing planet (at a distance to resonance $\delta_n=0.2n_{res}$). }
    \label{Pf9}
     \end{figure}

    It is important that the frequencies does not depend on the mass of the planet. Instead, the amplitude increases linearly with increasing mass of the planet (Fig. \ref{Pf9}). Note that the amplitude in Fig. \ref{Pf9} is close to the separatrix: it rapidly decreases with distance to the resonance, see Fig.10 in \cite{RP7}. At the same time, the role of the second frequency ($\Lambda$) increases with increasing mass of the disturbing planet: with a Jupiter as a perturber, we have a disturbance with almost the single frequency.
    
    Therefore, the nature of the frequency $\nu$ is clear: it is the orbital frequency or the mean motion $n$ (in the non-averaged case) or the resonance-related frequency (see equations (\ref{E22}, \ref{E27}). The other frequency ($\Lambda$) depends on the distance to resonance (by a complex way). The bounded solution of (\ref{E26}) corresponds to the imaginary value of $\Lambda$. After the substitution: $\omega_0=\nu/2+\delta$, where $\delta$ is the distance to the exact resonance (in mean motion), we obtain the following dependence for frequency: $\Lambda(\delta)$:
     \begin{equation}
     \Lambda=\sqrt{h_j^2\left(\frac{\nu^2}{64}+\frac{\nu}{8}\delta+\frac{3}{8}\delta^2+\frac{1}{2\nu}\delta^3+\frac{1}{4\nu^2}\delta^4\right)-\delta^2}.
     \end{equation}
     
\begin{figure}
       \centering 
         \includegraphics[width=8.7cm]{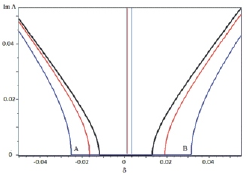}
         \caption {{The frequency of the asymptotic solution of the equation (\ref{E26}) in arbitrary units at different values of $h_j=0.01$ (black), $h_j=0.02$ (red), $h_j=0.05$ (blue). The vertical lines indicate the positions of the center of the unstable zones for each case. (see text for a details)} }
    \label{Pf15}
     \end{figure}

Denote A and B as the point of the dependence $Im(\Lambda)$ cross x-axis (i.e. $Im \Lambda=0 $). In the interval between points A and B in (Fig.\ref{Pf15}) the value of $\Lambda$ is real, so there is no bounded solution. This can be interpreted as follows: the solution is unstable in this interval, or our linear model is incomplete. 
     
Therefore, we can draw the following conclusion: the frequency of the solution of the linearized Mathieu equation decreases in approach to resonance, similarly as in the results of numerical integration.   
  
  This conclusion is a justification of the results obtained (\cite{R2}) in the special case of the resonance of three bodies. 
    
 \subsection{Comparison with the pendulum model}

 The solution of the pendulum equation can be performed in elliptic functions. According to \cite{W}, the solution of the pendulum Hamiltonian equation for case circulation is expressed through an elliptic integral. Compared with our model, it is possible to use an asymptotic solution. According to (\cite{Mi}):
\begin{equation}
    \delta \lambda=\delta_n t+const/\delta_n^2\sin \delta_nt,
\end{equation}
where $\delta_n$ is the perturbation frequency, in our case it coincides with the distance to the resonance in the mean motion, $\delta_n=n-n_{res}$.

 As is evident from the equation above, the frequency of perturbations tends to zero in approach to the separatrix (to the exact resonance). The amplitude of perturbations tends to infinity. 
 
In our linearized model, the amplitude close to the separatrix tends to infinity, but the frequency always has a non-zero value. This can be important to study the process of capture in resonance. 

  \subsection{Resonance instability}
 
 In a non-averaged problem, regions of parametric instability may arise. Assume that the unperturbed value of $\omega$ is close to the resonant mean motion:
  
 \begin{equation}
 \omega_0=\sqrt{G M/r_0^3}+\Delta=n_{res}+\delta 
 \end{equation} 
 
 
  where $\delta$ is the distance to the exact resonance (in mean motion). 
 The condition of the parametric resonance is:
 \begin{equation}
 \nu=2\omega/K=jl/(k+l)n
 \end{equation}
 where n is the mean motion, j, l, k, K are integers. The parametric resonance has notable width only when K=1 or K=2. 
 At a small eccentricity (e=0.1) it is possible to restrict ourselves by the first term of the equation \ref{E14}:
 \begin{equation}
  \cos j\delta\lambda\approx \cos j(\lambda-\lambda_p)=\cos j(n-n_p)t.
  \label{E00}
  \end{equation} 
  
  For the 3:2 resonance the equation is:
  \begin{equation}
       \ddot x+\omega^2_0(1-h_j\cos\omega_0 t)x=f(t)
       \label{E38}
      \end{equation}
 where (j=3):
    \begin{equation}
     h_j=h_3 =\frac {m r_{0}^3}{M R^3}b_{1/2}^3(\alpha).
      \label{E46}
      \end{equation}
     
  For the 2:1 resonance we need to account for the first and second parametric resonances at j=1 and j=2.
  Therefore in addition to equation (\ref{E38}) it is necessary to consider the equation (j=1):
 \begin{equation}
       \ddot x+\omega^2_0(1-h_j\cos\omega_0/2 t)x=f(t)
       \label{E39}
      \end{equation}
  	\begin{equation}
       h_j=h_1 =\frac {m r_{0}^3}{M R^3}b_{1/2}^1(\alpha),
       \end{equation}
    and equation (\ref{E46}) with j=2
       \begin{equation}
    h_j=h_2 =\frac {m r_{0}^3}{M R^3}b_{1/2}^2(\alpha),  
        \end{equation}
 The width of the unstable area is determined by the coefficient $b_j$ (\cite{Lf}):
\begin{equation}
 -\frac{m r_0^3}{2MR^3}b_j \omega_0 <\Delta<\frac{m r_0^3}{2MR^3}b_j \omega_0.  
  \label{EC}
 \end{equation}
 The width of the unstable zones for different resonances can be estimated in a similar way. Our results for some of the main resonances are given in Table 1. There are resonance notation in the first column, the value of j in the second column, $\alpha=r_0/R$ in the third column, value of the corresponding Laplace coefficient $b_{1/2}^{j}$ in the fourth column, the value of $h_j$ in the fifth column and the value of the width of unstable zone in the semimajor axis in the last column.  However, the position of the unstable zone can depend on eccentricity (\cite{R1}). 

In addition, a small offset of the center of the unstable zone relative to the resonance center is predicted \citep{Lf}. The offset may be important in studying the process of the interaction of the small body with resonance.

\begin{table*}[t]
\caption{Width of the unstable zones for different resonances}
\begin{center}
\begin{tabular}{l l r r r r r r} 
\hline 
\multicolumn{1}{| c }{Resonance} & \multicolumn{1}{l |}{}   & \multicolumn{1}{c |}{j}  & \multicolumn{1}{c |} {$\alpha$}  & \multicolumn{1}{c |} {$b_{1/2}^{(j)}$}   &     \multicolumn{1}{c |}{$h_j$}    &     \multicolumn{1}{c |}{$\delta_a$, AU}\\
\hline
\multicolumn{1}{| c }{2:1} & \multicolumn{1}{l |}{}  & \multicolumn{1}{c |}{2}  & \multicolumn{1}{c |} {0.629961}  & \multicolumn{1}{c |} {0.36531427}   &     \multicolumn{1}{c |}{0.091}    &     \multicolumn{1}{c |}{0.000199}\\
\multicolumn{1}{| c }{2:1} & \multicolumn{1}{l |}{}  & \multicolumn{1}{c |}{1}  & \multicolumn{1}{c |} {0.629961}  & \multicolumn{1}{c |} {0.75684039}   &     \multicolumn{1}{c |}{0.189}    &     \multicolumn{1}{c |}{0.000413}\\
\multicolumn{1}{| c }{3:2} & \multicolumn{1}{l |}{}  & \multicolumn{1}{c |}{3}  & \multicolumn{1}{c |} {0.763143}  & \multicolumn{1}{c |} {0.40051639}   &     \multicolumn{1}{c |}{0.178}    &     \multicolumn{1}{c |}{0.000471}\\
\multicolumn{1}{| c }{3:1} & \multicolumn{1}{l |}{}  & \multicolumn{1}{c |}{3}  & \multicolumn{1}{c |} {0.480750}  & \multicolumn{1}{c |} {0.07780038}   &     \multicolumn{1}{c |}{0.009}    &     \multicolumn{1}{c |}{0.000015}\\
\multicolumn{1}{| c }{5:3} & \multicolumn{1}{l |}{}  & \multicolumn{1}{c |}{5}  & \multicolumn{1}{c |} {0.711378}  & \multicolumn{1}{c |} {0.12285507}   &     \multicolumn{1}{c |}{0.044}    &     \multicolumn{1}{c |}{0.000109}\\
\multicolumn{1}{| c }{5:2} & \multicolumn{1}{l |}{}  & \multicolumn{1}{c |}{5}  & \multicolumn{1}{c |} {0.542884}  & \multicolumn{1}{c |} {0.02718458}   &     \multicolumn{1}{c |}{0.004}    &     \multicolumn{1}{c |}{0.000007}\\
\multicolumn{1}{| c }{7:4} & \multicolumn{1}{l |}{}  & \multicolumn{1}{c |}{7}  & \multicolumn{1}{c |} {0.688612}  & \multicolumn{1}{c |} {0.04133825}   &     \multicolumn{1}{c |}{0.014}    &     \multicolumn{1}{c |}{0.000033}\\
 \hline 
\end{tabular}
\end{center}
\label{tab1}
\end{table*}

It does not mean that orbits near the low order resonances are instable parametrically. The width of the unstable zone is much smaller than the width of a first order resonance. This means that only the presence of an unstable zone at the boundary of the resonance corresponded with the separatrix between the libration and the circulation motion.

However, in the case of weak resonances, the width of the instability zone can be wider than the width of the resonance. All resonances with Mars and Earth in the Main Belt may be sufficiently weak. In the case of such weak resonances, capture in the libration regime is very unlikely or impossible, and our model can be useful in studying the transfer of such resonances. 

\section{Verification of the model}

To test our theoretical model, we integrate the elliptical three body problem (Sun-Jupiter-Asteroid). In this paper, we present our results of numerical integration of the orbits of 23 fictive massless particles near resonance with an eccentricity of 0.1 and a step along the initial semimajor axis equal to 0.005 AU. 
The massless test particle (asteroid) orbited in a different distance $d $,  close to the resonance with Jupiter (i.e. with semi-major axis $a_{res}+d$).  However, the average values of the semimajor axes are notably different from the initial ones (due to the planet perturbation). All orbits start at the perihelion which coincides with the planet perihelion. Two series of integrations were performed with the circular orbit of Jupiter and with the elliptical one.
 We used a standard package Mercury in our integration (\cite{Ch1}, \cite{E1}). 
 For more details of our integration see the preprint \cite{RP7}. It can be considered as supplementary material for this paper. 

In order to make a comparison with the results of numerical integration, it is necessary to transfer from changes in the radius vector to the evolution of the elements of the orbits (in the planar case $a$ and $e$). The relationship between $a$ and $e$ is given by equation (8.35) from the book by \cite{MD}. This means that the frequency (period) of the resonant disturbance is the same for the semimajor axis and eccentricity as follows for the radial distance from central mass.


Therefore, we can use the changes in the radial distance from the central mass \textcolor{blue}{$x=r-r_0$, see equation (\ref{E6}) } as the measure of the perturbations in the orbital element to describe the resonance. The same conclusion follows from the Lagrange equations.


For quantitative estimation, the eccentricity and the semimajor axis of the obtained orbits were approximated by a trigonometric polynomial by our method \cite{RP5}.

To obtain the approximation, we employ a minimum of the following function:

\begin{equation}
  \sigma=\sqrt{\frac{1}{n-2} \sum\limits_{i=1}^{n}(X_i-X_{iapprox})^2}
  \end{equation}

Here $n$ is the number of points, $X_i$ is the value obtained by the numerical solution,  $X_{iapprox}$ is the approximating value and the result is confirmed by Fourier approximations.
Finally, we approximate the obtained values of the frequency by straight lines (separating above and below the exact resonance) using standard software (Excel for Windows).

We made series of integrations with different disturbing planet masses (see \cite{RP7}). First, we assume that the eccentricity of Jupiter is equal to zero.  In the results we have obtained,  the dependence of the frequency of perturbation on the distance to resonance is in close agreement with our model (see Fig.8 and Fig. 9 in \cite{RP7}). The frequency of perturbation decreases in approach to the resonance as is shown in Fig.\ref{Pf4}.  After that, we repeated our integrations and approximation with the actual eccentricity of Jupiter (e=0.048) and obtained the same values of frequency and amplitude.
  
  \subsection{Case 3:2 resonance}

This subsection was previously published in the preprint of \cite{RP7}. The value of the semimajor axis corresponded with the exact commensurability with Jupiter $a_{res}$ is slightly different by different authors. By the \cite{MD} book $a_{res} = 3.970320$ AU, by the \cite{SD} paper $a_{res} = 3.96897$ AU. By our previous numeric integrations, the value, averaged over 800 kyr interval is $a_{res} = 3.970319$ AU is very close to the \cite{MD} value.
The multiplet term, corresponding to the Jupiter perihelion rotation, $a_{e} = 3.97042$ AU is very close to the nominal resonance position.

At the outer boundary, the resonance 3:2J with Jupiter can overlap with the 4:3J resonance (centered at about 4.2946415 AU). The rough estimation of the boundary is at about 4.132481 AU. To avoid this, we use the reduced mass of the perturbing planet (Jupiter) to the detailed study the 3:2 resonance. To exclude secular perturbations, we assume that the eccentricity of the Jupiter orbit is equal zero. We integrate several fictive asteroids with different initial values of the semimajor axis.

As it is known, there are two kinds of motion close to resonance with the libration and the circulation of the resonant argument. Here we point the main attention to the study of the circulation solution. In result we obtain the periodic perturbations in the semimajor axis.  A similar behaviour takes place for the asteroid eccentricity.

Then we quantitatively determine the main amplitudes and frequencies of perturbations in the semimajor axis. The obtained values are confirmed by the Fourier analyses method. The frequency of the perturbations increases with distance to the exact resonance (see Fig. \ref{Pf4}) when the amplitude decreases. Finally, we approximate the obtained values of the frequency by straight lines (separating above and below the exact resonance). 

To study the behavior of the frequency of the numerical solution depending on the mass of the planet, we conducted a test for a system (Sun-Planet-asteroid) with a planet mass = 0.01 $m_J$, 0.1 $m_J$, 0.5 $m_J$.
The results are the following:

For mass planet $m=0.01m_J$
\begin{equation}
f=-593.16a + 2355.6 \  
f=570.62a - 2264.9.          
\end{equation}
The corresponding intersection point ordinate is $a=3.9702521$ AU.
For mass planet $m=0.1m_J$
\begin{equation}
f=-550.38 a + 2188.1\   
f=607.86 a - 2414.             
\end{equation}
The corresponding intersection point ordinate is $a=3.973356$ AU.
 For mass planet $m=0.5m_J$
 \begin{equation}
 f=-476.93 a+ 1900.2\  
 f=555.3a - 2202.7.             
 \end{equation}
 The corresponding intersection point ordinate is $a=3.974792$ AU. 

 We believe that the position of the intersection point determines the position of the resonance. As a result, we note, that the position of resonance moved from nominal in a direction of larger semimajor axis when the perturbing planet mass increases. In the same time, the value of the frequency at the intersection point increases with the planet mass. The amplitude of the semimajor axis variations increase linearly with increasing mass of the planet (as it is shown in Fig. \ref{Pf9}).
          
\section{Discussion and conclusions}

A method of linearization of the equation of motion close to (but not exactly in) resonance  is proposed under the assumption of small perturbations. In the other words, we study orbits when the resonant argument circulates around the resonance. In the practically important case of resonant perturbations, the analysis can be limited to the dominant term with a single frequency of largest amplitude, reducing the problem to the Mathieu equation.

	First, we reduced the equations of motion to the Hill and Mathieu (in the case close to the resonance) equations. The resulting expressions were derived under the assumption of small perturbations, $x<<R$. 
 
Then, we study the properties of the asymptotic solution of the obtained Mathieu equation and compare it with the motion of a real asteroid in the vicinity of resonance in the three-body problem.
Finally, we provide the qualitative comparison of our analytic model with some important attributes of resonant motion. This is a useful preface before proceeding to quantitative analysis, because some features of resonant motion still have less attention than necessary. 

As a result, we conclude that our model can explain (at least qualitatively) the following attributes of resonant motion:

1.	the presence of perturbations with two different frequencies in semimajor axis and eccentricity even in the RTBP case;

2.	the existence of a parametrically unstable zone near the resonance;

3.	the dependence of the resonant perturbation frequency and amplitude on the distance to the resonance.

4.	offset of the resonance center from its nominal position in the case of a large disturbance force (mass of the perturbing planet);

Because the model is linear, it properly accounts for resonant perturbations outside the separatrix, that is,in the circulation regime. This can be useful for a wide range of applications, since circulation-type motion occurs far more frequently than exact resonance libration. The study of libration, however, requires a nonlinear approach such as, for example, a pendulum model. Although a narrow instability zone persists near the separatrix, our model correctly identifies its location and thus the transition boundary between libration and circulation of the resonant argument.

\section{conflicts of interest}
The authors declare that they have no conflicts of interest

\section{ Ethics declaration}
 The authors declare that present paper is written in agreement with Committee on Publication Ethics requirements.
 
\begin{acknowledgements}
The authors kindly acknowledging the associate editor and the reviewers in the paper, given their informative and helpful comments. The contribution of the third author (P. Nesterov) was carried out within the framework of a development programme for the Regional Scientific and Educational Mathematical Center of the Yaroslavl State University with financial support from the Ministry of Science and Higher Education of the Russian Federation (Agreement on provision of subsidy from the federal budget No. 075-02-2025-1636) 
\end{acknowledgements}

\section{Statements $\&$ Declarations}
The authors declare that no funds, grants, or other support were received during the preparation of this manuscript

\end{document}